\documentclass[prl,superscriptaddress,twocolumn,amsmath,amssymb,nofootinbib,longbibliography]{revtex4-2}

\usepackage{graphicx}
\usepackage{amsmath}
\usepackage{physics}
\usepackage{color}
\usepackage{xcolor}
\usepackage{slashed}
\usepackage{tensor}
\usepackage[utf8]{inputenc}
\usepackage[normalem]{ulem}

\newcommand{\ba}{\begin{eqnarray}}
\newcommand{\ea}{\end{eqnarray}}

\newcommand{\be}{\begin{equation}}             %:skip:
\newcommand{\ee}{\end{equation}}               %:skip:
\newcommand{\notprop}{\propto\kern-1\@ptsize pt \diagup}

\begin{document}

\title{Thermodynamically Stable Phases of Asymptotically Flat
Lovelock Black Holes}

\author{Jerry Wu}
\email{yq4wu@uwaterloo.ca} 
	\affiliation{Department of Physics and Astronomy, University of Waterloo,
		Waterloo, Ontario, Canada, N2L 3G1}
	
\author{Robert B. Mann}
	\email{rbmann@uwaterloo.ca}
		\affiliation{Department of Physics and Astronomy, University of Waterloo,
		Waterloo, Ontario, Canada, N2L 3G1}	
	\affiliation{Perimeter Institute, 31 Caroline Street North, Waterloo, ON, N2L 2Y5, Canada}

\date{\today}
	
	\begin{abstract}
		We present the first examples of phase transitions in asymptotically flat black hole solutions. We analyze the thermodynamic properties of black holes in order $N\ge 3$ Lovelock gravity, with zero cosmological constant. We find a new type of ``inverted'' swallowtail indicative of stable temperature regions for an otherwise unstable neutral black hole, and demonstrate multiple such stable phases can exist and coexist at multi-critical points. We also find that for charged black holes, ordinary swallowtails can exist on the stable Gibbs free energy branch, allowing for multiple first order phase transitions as seen for AdS black holes. A triple point for $N=5$ and a quadruple point for $N=7$ are presented explicitly. We investigate changes in the Gibbs free energy as the lowest order  Lovelock constant is varied, and draw comparisons to pressure changes for AdS black hole systems.
	\end{abstract}

\maketitle

\section{introduction}

Black hole thermodynamics has played a vital role in providing clues to a full description of quantum gravity. Previously thought to be perfect absorbers, black holes have been shown to emit thermal radiation with a temperature proportional to the surface gravity and an entropy proportional to the horizon area after taking quantum effects into account \cite{Hawking:1975vcx}. Further studies on asymptotically anti de Sitter (AdS) black holes by Hawking and Page revealed that a first order phase transition was possible between a large Schwarzschild-AdS black hole and   thermal AdS \cite{Hawking:1982dh}. This has become known as the Hawking-Page (HP) transition.

In general, the negative cosmological constant in AdS space can be interpreted as  thermodynamic pressure
\cite{Kubiznak:2016qmn}, and the black hole mass takes on the role of enthalpy in this context \cite{Kastor:2009wy}. With this realization, a plethora of thermodynamic phenomenon was observed in AdS black holes, including reentrant transitions \cite{Altamirano:2013ane}, superfluid black holes \cite{Hennigar:2016xwd}, snapping transitions \cite{Abbasvandi:2018vsh}, and triple points \cite{Altamirano:2013uqa,Frassino:2014pha,Wei:2014hba,Wei:2021krr}. Very recently  multi-critical points have been shown to exist.  These occur in 4-dimensional Einstein gravity coupled to power Maxwell theory \cite{Tavakoli:2022kmo}, and for multiply rotating Kerr-AdS black holes \cite{Wu:2022bdk}. In general, it appears that pretty much any phase behaviour observed in a chemistry lab can also appear in an AdS black hole system, and for this reason the subject has come to be called Black Hole Chemistry \cite{Kubiznak:2016qmn}.

It is also expected that to account for quantum gravitational effects, higher order curvature corrections  to the Einstein-Hilbert action 
will appear \cite{Birrell:1982ix}. The most physically significant such generalization is Lovelock gravity \cite{Lovelock:1971yv,Lovelock:1972vz}, which has field equations that are second order in the metric. For these reasons, the study of Lovelock black hole thermodynamics has become of substantial interest, despite requiring higher spacetime dimensions to produce novel results. Comprehensive studies of Lovelock AdS black holes 
(and other theories incorporating higher curvature effects \cite{Hendi:2018xuy, Dehghani:2020blz,Mir:2019ecg,Mir:2019rik,Bueno:2022res})
have shown that they exhibit a   rich variety of thermodynamic behaviour, such as reentrant phase transitions \cite{Frassino:2014pha}, triple points \cite{Frassino:2014pha,Wei:2014hba,Hull:2021bry,Wei:2021krr,Hull:2022xew}, isolated critical points \cite{Frassino:2014pha,Dolan:2014vba}, and most recently, multi-critical points in higher order Lovelock theories \cite{Wu:2022plw}. The zeroth law of black hole thermodynamics has also been studied in higher curvature gravity \cite{Ghosh:2020dkk}.

Asymptotically flat black holes in Einstein gravity have never been observed to undergo first order phase transitions. The Schwarzschild black hole is locally thermodynamically unstable and no phase transitions are seen in the Gibbs free energy. The Reissner–Nordstr\"{o}m black hole has one stable branch with no phase transitions in the Gibbs free energy, and the Kerr black hole has similar properties \cite{Altamirano:2014tva}. Furthermore, 
 very few investigations 
in  Lovelock gravity
have focused on the thermodynamics of   asymptotically flat black holes \cite{Ma:2016vop,Sinha:2021dlm}, with studies  limited to second-order (Gauss-Bonnet) and  third-order Lovelock theories \cite{Dehghani:2005vh}, where a stable phase was shown to exist
for a range of horizon radii between some minimal and maximal values.

Here we investigate the phase behaviour of 
uncharged and charged asymptotically flat black holes in  Lovelock gravity. 
We find that such black holes can in fact undergo multiple phase transitions and exhibit multicritical behaviour when higher curvature terms are taken into consideration. The Gibbs free energies of these two classes of black holes largely resemble their  counterparts in Einstein gravity. However unlike Einstein gravity,  multiple zeroth and first order phase transitions are now possible on both stable and unstable branches with a sufficient but finite number of Lovelock coupling constants. These  coupling constants can be regarded as thermodynamic variables \cite{Frassino:2014pha,Kastor:2010gq,Kastor:2011qp}, and with the realization that the number of coexistence phases  increases with the number of thermodynamic variables \cite{Sun:2021gpr}, we investigate their multicritical behaviour. We find that indeed
such phase transitions can merge at multi-critical points and terminate at critical points, as observed in AdS black holes. An uncharged black hole can support mulitple thermodynamically stable phases, generalizing the situation for third-order Lovelock gravity 
\cite{Dehghani:2005vh}.
Although no thermodynamic pressure is present, we find that the Gauss-Bonnet coupling constant plays a role similar to pressure on the phase behaviour of black hole systems. Novel phase behaviour, such as ``inverted'' swallowtails, maximal temperatures, phase transitions below the multi-critical temperature, and multiple termination points are discussed.

We begin by reviewing the basic thermodynamics of Lovelock black holes. We then discuss the phase behaviour of uncharged asymptotically flat black holes, and illustrate their phase behaviour and multicriticality in several examples. In section \ref{sect4} we consider the charged case, and show that there is a thermodynamically stable branch of charged black holes that can have multiple phases and multicritical behaviour.
We sum up our results in a concluding section.

In general our results indicate that asymptotically flat black holes can exhibit phase behaviour similar to their AdS counterparts, provided the number of thermodynamic parameters is sufficiently large and that the temperature has a sufficiently non-linear dependence on the horizon radius.

\section{asymptotically flat lovelock solutions}

An $N$th-order Lovelock theory of gravity in $d$ spacetime dimensions is described by the Lagrangian \cite{Lovelock:1971yv}
\be
\mathcal{L}=\frac{1}{16 \pi G_N}\sum_{k=0}^N \hat{\alpha}_k \mathcal{L}^{(k)} \label{lagrangian}
\ee
with $\hat{\alpha}_k$ being the $k$th order Lovelock coupling constants, and $\mathcal{L}^{(k)}$ being the $2k$-dimensional Euler densities
\be
\mathcal{L}^{(k)}= \frac{1}{2^k} \delta_{c_1 d_1 \dots c_k d_k}^{a_1 b_1 \dots c_k b_k} R_{a_1 b_1}^{\quad c_1 d_1} \dots R_{a_k b_k}^{\quad c_k d_k}.
\ee
The terms $\mathcal{L}^{(0)}$, $\mathcal{L}^{(1)}$, and $\mathcal{L}^{(2)}$ coincide with the cosmological constant term, Einstein-Hilbert term, and Gauss-Bonnet term in the Lagrangian respectively.

Typically, $\hat{\alpha}_0$ is interpreted as a thermodynamic pressure induced by a negative cosmological constant
\be
\hat{\alpha}_0 = - 2 \Lambda, \qquad P=-\frac{\Lambda}{8 \pi G_N} = \frac{\hat{\alpha}_0}{16 \pi G_N}
\ee
where $G_N$ represents Newton's constant. To obtain asymptotically flat solutions, we set $P=\hat{\alpha}_0=0$. We also require $d\ge 2N+1$ as the higher order Euler density term $\mathcal{L}^{(N)}$ does not make a contribution to the equations of motion for $d\le 2N$.

A charged black hole with electromagnetic field $F=dA$ has action
\be
I=\int d^d x \frac{\sqrt{-g}}{16 \pi G_N} \left(\sum_{k=1}^N \hat{\alpha}_k \mathcal{L}^{(k)} - 4 \pi G_N F_{ab} F^{ab}\right)
\ee
as follows from \eqref{lagrangian}, and equations of motion
\be
\sum_{k=1}^N \hat{\alpha}_k \mathcal{G}_{ab}^{(k)} = 8\pi G_N \left(F_{ac} F_b^{\ c} - \frac{1}{4} g_{ab} F_{cd} F^{cd} \right), \label{fieldeq}
\ee
where the Einstein-like tensors $\mathcal{G}^{(k)}$ are of the form
\be
\mathcal{G}^{(k) a}_{\ b} = -\frac{1}{2^{k+1}} \delta^{a c_1 d_1 \dots c_k d_k}_{b e_1 f_1 \dots e_k f_k} R_{c_1 d_1} ^{\quad e_1 f_1} \dots R_{c_k d_k} ^{\quad e_k f_k},
\ee
and each satisfies the conservation law $\nabla _a \mathcal{G}^{(k) a}_{\ b}=0$ independently.

The metric ansatz used for static spherically symmetric black holes with charge $Q$ is as follows
\begin{align}
    ds^2 &= -f(r) dt^2 + f(r)^{-1} dr^2 + r^2 d\Omega^2 _{(\kappa)d-2}, \nonumber \\
    F &= \frac{Q}{r^{d-2}} dt \wedge dr,
\end{align}
which reduces \eqref{fieldeq} to an $N$th-degree polynomial equation in the metric function $f$ \cite{Boulware:1985wk,Wheeler:1985nh,Wheeler:1985qd,Cai:2003kt,Camanho:2011rj,Takahashi:2011du,Castro:2013pqa}:
\begin{align}\label{lovepoly}
\sum_{k=1}^{N} \alpha_k & \left(\frac{\kappa -f }{r^2} \right)^k  = \Upsilon(r; M,Q)
\end{align}
where
$$
\Upsilon(r; M,Q) = \frac{16 \pi G_N M}{(d-2)\Sigma^{(\kappa)}_{d-2}r^{d-1}} - \frac{8 \pi G_N Q^2}{(d-2)(d-3)r^{2d-4}}
$$
and the $\alpha_k$ are the rescaled Lovelock constants
\begin{equation} \alpha_1=\hat{\alpha}_1, \qquad
\alpha_k = \hat{\alpha}_k \prod_{n=3}^{2k} (d-n) \text{ for } k\ge2 \; .
\end{equation}
We shall limit ourselves to consider only spherical horizon geometries, for which $\kappa=+1$, and 
\be
\Sigma^{(+1)}_{d-2} = \frac{2\pi ^{(d-1)/2}}{\Gamma(\frac{d-1}{2})}.
\ee

It is not possible to obtain an analytic solution to \eqref{lovepoly} for $N>4$.  However we can obtain a series solution using the ansatz 
\be\label{fansatz}
    f(r) = 1 + r^2\sum_{i=1}^{\infty}  \zeta_i (r; M, Q),
\ee
which solves  \eqref{lovepoly} provided  
\begin{align}
    \zeta_1 &= -\Upsilon, \quad \zeta_2 = \Upsilon^2 \alpha_2, \quad \zeta_3 = -\Upsilon^3 (2\alpha_2^2 - \alpha_3) \nonumber \\ 
    \zeta_4&=\Upsilon^4 (5 \alpha_2^3 - 5 \alpha_2 \alpha_3 + \alpha_4) \nonumber \\
    \zeta_5&=-\Upsilon^5 (14 \alpha_2^4 - 21 \alpha_2^2 \alpha_3 + 6 \alpha_2 \alpha_4 +3 \alpha_3^2 - \alpha_5) \nonumber \\ 
    \zeta_6&=\Upsilon^6 (42 \alpha_2^5 - 84 \alpha_2^3 \alpha_3 + 28 \alpha_2^2 \alpha_4 \nonumber \\ & \qquad + 28 \alpha_2 \alpha_3^2 - 7 \alpha_2 \alpha_5 - 7\alpha_3 \alpha_4 + \alpha_6) \nonumber \\
    \dots &= \dots \dots
\end{align}
using series reversion.   
The solution is asymptotically flat as $\lim_{r\to \infty} f(r)=1$  and for any given value of $N$ the
coefficients in the
infinite series \eqref{fansatz}   uniquely depend on the $N-1$ independent coefficients  $(\alpha_2,\alpha_3,\ldots,\alpha_{N})$.
As the $\alpha_i \to 0$ its limit is the electrovacuum solution to $d$-dimensional Einstein gravity.  This solution is known as the Einstein branch; there are $N-1$ other solutions to \eqref{lovepoly} corresponding to the other Lovelock branches.

Fortunately we need not solve
\eqref{lovepoly} explicitly for $f$, since it is possible to determine the thermodynamic characteristics of a black hole, including the mass $M$, temperature $T$, entropy $S$, and electromagnetic gauge potential $\Phi$, as functions of the horizon radius and $Q$
\cite{Cai:2003kt,Kastor:2011qp} in Planckian units 
$l^2_P =\frac{G\hbar}{c^3} $ \cite{Kubiznak:2016qmn}, 
\begin{align}
    M&=\frac{\Sigma^{(+1)}_{d-2}(d-2)}{16 \pi G_N}\sum_{k=1}^{N} \alpha_k r_+^{d-1-2k} + \frac{\Sigma^{(+1)}_{d-2} Q^2}{2(d-3)r_+^{d-3}},\\
    T&= \frac{1}{4\pi r_+ D}\left[ \sum_k \alpha_k \frac{d-2k-1}{r_+^{2(k-1)}} - \frac{8\pi G_N Q^2}{(d-2) r_+^{2(d-3)}}\right], \\
    S&= \frac{\Sigma^{(+1)}_{d-2} (d-2)}{4 G_N} \sum_{k=1}^{N} \frac{k \alpha_k r_+^{d-2k}}{d-2k},\\
    \Phi &= \frac{\Sigma^{(+1)}_{d-2} Q}{(d-3) r_+^{d-3}},
\end{align}
where $r_+$ is given by the largest root of $f(r_+)=0$, and $D$ is the series
\be
    D=\sum_{k=1}^N k \alpha_k r_+^{-2(k-1)}.
\ee

Thermodynamic quantities of a black hole are related by the extended first law of black hole thermodynamics \cite{Jacobson:1994qe} and the generalized Smarr relation \cite{Kastor:2010gq}
\be
\delta M = T \delta S - \frac{1}{16 \pi G_N} \sum_k \hat{\Psi}^{(k)} \delta \hat{\alpha}_k + \Phi \delta Q
\ee
\be
 M = \frac{d-2}{d-3} T S + \sum_k 2\frac{k-1}{d-3} \frac{\hat{\Psi}^{(k)} \hat{\alpha}_k}{16 \pi G_N} + \Phi Q.
\ee
where the potentials $\hat{\Psi}^{(k)}$ are thermodynamic conjugates of the Lovelock constants $\hat{\alpha}_k$ \cite{Kastor:2010gq}.

To study thermodynamic phase transitions, we consider Gibbs free energy 
\be
    G=M-TS=G(T,Q,\alpha_1,\dots,\alpha_N),
\ee
for the thermodynamically favourable state coincides with the global minimum of $G$.

We also fix $\alpha_1 = G_N = 1$ to recover General Relativity in the low curvature limit, and restrict all $\alpha_k$ to be strictly  nonnegative in order avoid nakedly singular solutions.

\section{Uncharged black holes}

 In order to observe multi-criticality, a sufficiently large number of phases is required, which in turn requires a sufficiently large number of Lovelock coupling constants $\alpha_n$.  In this section we shall set $N=7$ (so that there are five distinct couplings) in our examples, though the behaviour we describe is valid for any finite $N$. We shall also set $d=16$, which is the minimum even dimension at which the $N=7$
Lovelock action is non-trivial.

\begin{figure}
	\includegraphics[width=0.483\textwidth]{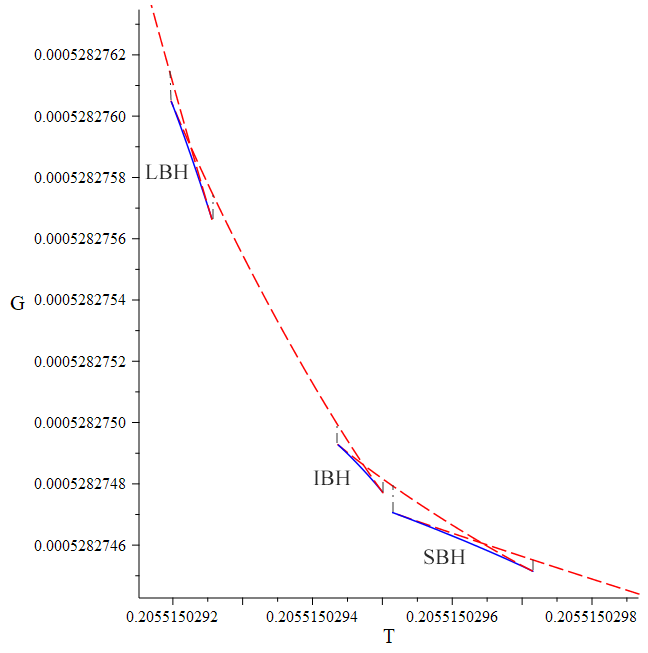}

	\caption{\textbf{$G$-$T$ plot: Three separated inverted swallowtails.} Here $d=16$, $N=7$, $\alpha_2 = 7.5241184$, $\alpha_3 \approx 16.618660$, $\alpha_4 \approx 12.8004276$, $\alpha_5 \approx 3.5841701$, $\alpha_6 \approx 0.325566891$, $\alpha_7 \approx 0.0064404661$. Three inverted swallowtails, each supporting a stable phase, are completely separated. Dashed red curves indicate negative specific heat. Dashed gray lines represent zeroth order phase transitions.}
	\label{fig:7ll-gt-separated} 
\end{figure}

Without a nonzero cosmological constant playing the role of thermodynamic pressure,    the phase behaviour is governed by  the next smallest unfixed coupling parameter, which we will take to be $\alpha_2$. Changes in $\alpha_2$ function similar to changes in pressure, insofar as the presence and distribution of swallowtails in the $G$ vs. $T$ diagram is affected.  Qualitatively, for small $\alpha_2$, the Gibbs free energy  has an unstable branch with negative heat capacity, resembling that of the Schwarzschild black hole \cite{Altamirano:2014tva}, with $G$ a monotonically decreasing function of $T$.   
 These black holes will become hotter as they emit Hawking radiation, until they reach a size where the semiclassical approximation breaks down.   
 As $\alpha_2$ increases, an inverted swallowtail appears along the curve, with the lower part of the swallowtail corresponding to a range of stable black holes with positive heat capacity.  As $\alpha_2$ further increases, more such swallowtails appear, until
eventually a set of  $\lfloor \frac{N-1}{2}\rfloor $  stable black holes is present over a discrete range of temperatures. In figure~\ref{fig:7ll-gt-separated} for
$N=7$ and $d=16$ we illustrate  three discrete (i.e. non-overlapping) inverted swallowtails, the maximal number allowed for this value of $N$
and the minimum number required to obtain a quadruple point.

  A black hole created in these stable thermodynamic regions (by whatever mechanism) will cool off as it evaporates to smaller size, whereas one created in the unstable regions will heat up as it evaporates. We see in figure~\ref{fig:7ll-gt-separated}  that there are 6 zeroth-order transition regions (two for each stable black hole), but only the three at the coldest respective temperatures are realizable without contrivance.  Consider, for example, a black hole created in the unstable region between a large black hole (LBH) and an intermediate black hole (IBH). This black hole will have increasing temperature as it evaporates. The system will move rightward along the dashed curve until the zeroth-order transition with the IBH occurs, at which point the size of the black hole will suddenly decrease to be the smallest allowed IBH.  This IBH will not evaporate as it is the smallest allowed black hole within this stable class. Similar remarks apply for the LBH and the small black hole (SBH):  an unstable black hole that is larger than either of these respective classes will evaporate until it undergoes a transition to the smallest black hole in these respective classes, at which point no further evaporation can take place.

\begin{figure}
	\includegraphics[width=0.483\textwidth]{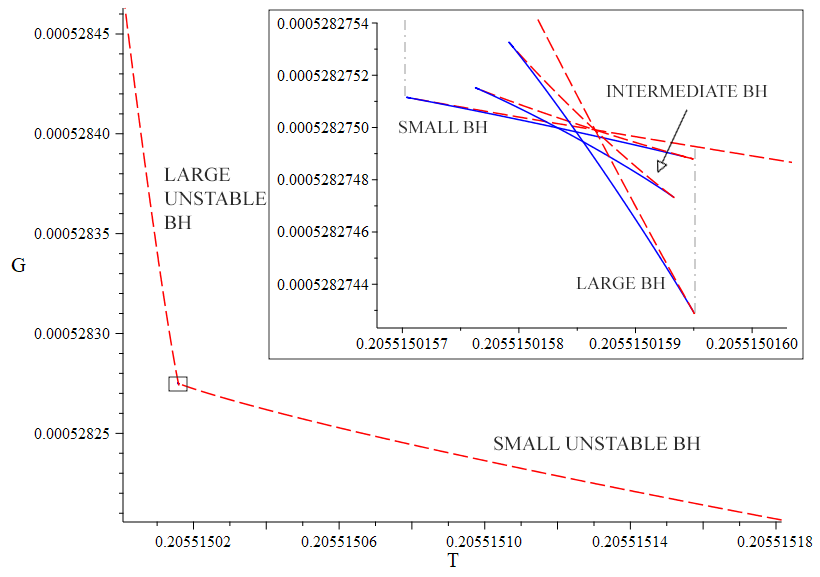}
	\caption{\textbf{$G$-$T$ plot: Unstable quadruple point of a neutral black hole.} Here $d=16$, $N=7$, $\alpha_2 \approx 7.524120$, $\alpha_3 \approx 16.618660$, $\alpha_4 \approx 12.8004276$, $\alpha_5 \approx 3.5841701$, $\alpha_6 \approx 0.325566891$, $\alpha_7 \approx 0.0064404661$. Three reversed swallowtails in the Gibbs free energy merge at a quadruple point between four unstable black hole phases. Dashed red curves indicate negative specific heat. Dashed gray lines represent zeroth order phase transitions.}
	\label{fig:7ll-gt-unstable-quad} 
\end{figure}

 Depending on the values of the other couplings, multicritical behaviour becomes possible along the unstable branch for any fixed value of $N$
 due to the added Lovelock coupling constants.
 Unlike ordinary swallowtails, inverted swallowtails merge at unstable multi-critical points, as the Gibbs free energy does not attain the global minimum value at the point of coexistence.
  Figure~\ref{fig:7ll-gt-unstable-quad} shows, for $N=7$ and $d=16$,  a system with 3 overlapping inverted swallowtails,   each having a region of stability. We see that two distinct first order phase transitions take place in the stable region, along with two zeroth order transitions between the stable and unstable states. An LBH created with a temperature  in the stable region will cool as its evaporates, undergoing a first-order transition to an IBH, and then another to an SBH.  The smallest SBH will cease evaporation unless cooled further by some contrivance, in which case it will undergo a zeroth order phase transition to an unstable LBH; without further external contrivance, this LBH will evaporate until it returns via a zeroth order transition to the stable SBH.  Likewise, an LBH in the stable region that is sufficiently heated will undergo a zeroth-order transition to an unstable SBH.  In this case the SBH will continue to heat up as it evaporates, and will not return to the stable LBH phase.
   In figure~\ref{fig:7ll-gt-unstable-quad} the zeroth order phase transition from unstable LBH to stable SBH occurs at $T \approx 0.2055150157$. The other zeroth order phase transition (from stable LBH to unstable SBH) is at $T \approx 0.20551501595$. 

\begin{figure}
	\includegraphics[width=0.483\textwidth]{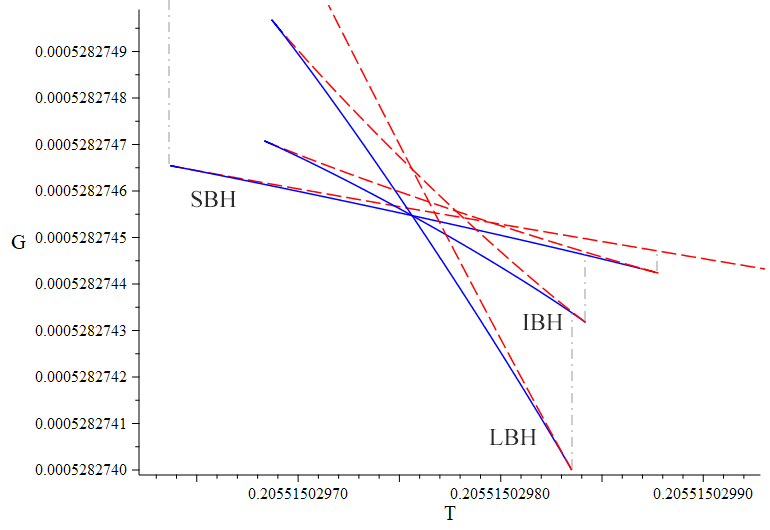}

	\caption{\textbf{$G$-$T$ plot: Stable triple point of a neutral black hole.} Here $d=16$, $N=7$, $\alpha_2 = 7.52411896$, $\alpha_3 \approx 16.618660$, $\alpha_4 \approx 12.8004276$, $\alpha_5 \approx 3.5841701$, $\alpha_6 \approx 0.325566891$, $\alpha_7 \approx 0.0064404661$. Three stable phases in the Gibbs free energy intersect at a triple point. Dashed red curves indicate negative specific heat. Dashed gray lines represent zeroth order phase transitions.}
	\label{fig:7ll-gt-stable-tri} 
\end{figure}

A curious consequence of the separation of inverted swallowtails is observed for $\alpha_2$ slightly below the unstable quadruple point; the three stable phases intersect at a single point, giving rise to a {\it stable} triple point, which we illustrate in figure~\ref{fig:7ll-gt-stable-tri}. 
 At this value of $\alpha_2$ there is still only one zeroth order phase transition on the colder end of the stable region, but due to the separation of the inverted swallowtails, three zeroth order phase transitions are seen at the hotter end. The most energetically favourable LBH state terminates
at a temperature where stable IBH and SBH phases exist. Likewise, the most energetically favourable IBH state terminates at a point where the stable SBH is still present.  An SBH created in this region will cool as it evaporates, undergoing a zeroth order transition to the stable IBH, which in turn has a similar transition to the stable LBH.  This LBH will evaporate until the stable triple point is reached; for temperatures below this the system returns to the stable SBH phase.  We therefore have an example of a reentrant phase transition combined with a triple point for asymptotically flat black holes.

\begin{figure}
	\includegraphics[width=0.483\textwidth]{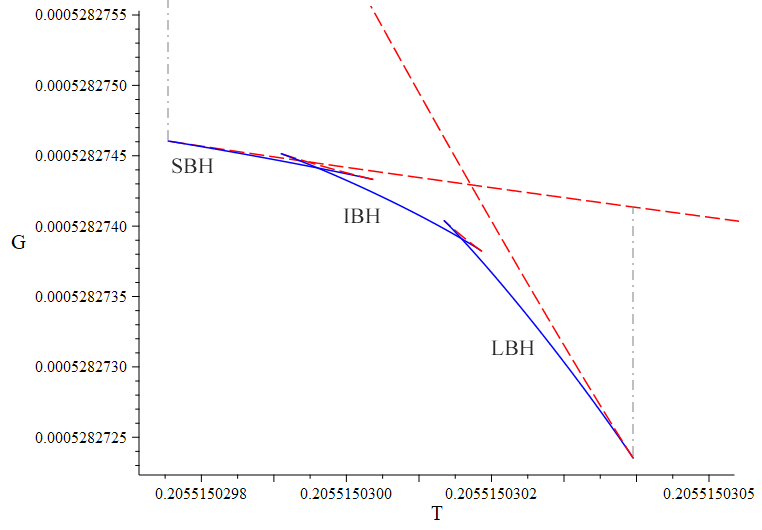}

	\caption{\textbf{$G$-$T$ plot: Embedded swallowtails.} Here $d=16$, $N=7$, $\alpha_2 = 7.5241195$, $\alpha_3 \approx 16.618660$, $\alpha_4 \approx 12.8004276$, $\alpha_5 \approx 3.5841701$, $\alpha_6 \approx 0.325566891$, $\alpha_7 \approx 0.0064404661$. Two ordinary swallowtails are seen on the stable branch of a large inverted swallowtail. Dashed red curves indicate negative specific heat. Dashed gray lines represent zeroth order phase transitions.}
	\label{fig:7ll-gt-embedded} 
\end{figure}

As $\alpha_2$ becomes greater than the multi-critical value, all swallowtails but one become embedded on the stable branch of a larger inverted swallowtail. We display an example of two embedded swallowtails for $d=16$ and $N=7$ at $\alpha_2=7.5241195$ in figure~\ref{fig:7ll-gt-embedded}. Given that they are positioned on a branch with positive heat capacity, the embedded swallowtails are now ordinary swallowtails, depicting first order phase transitions between three stable phases. Further increases in $\alpha_2$ yield behaviour that is qualitatively the same as for
asymptotically AdS black holes as the pressure increases. 
 The stable swallowtails separate and terminate at critical points, such that only one stable phase exists for sufficiently large $\alpha_2$. Zeroth order phase transitions are still present between the stable phase the unstable phases, and do not vanish for large $\alpha_2$. In fact, the region of stability expands as $\alpha_2$ increases. 

The number of thermodynamic variables required for $n=\lfloor \frac{N-1}{2} \rfloor$ coexisting phases is larger than that observed for previous systems \cite{Tavakoli:2022kmo,Wu:2022bdk}. The minimum thermodynamic degrees of freedom for this class of black holes using our methods is $n+2$ in accord with the generalized Gibbs phase rule
\cite{Sun:2021gpr} 
\be
\textsf{F}=\textsf{W}-\textsf{P}+1,
\ee
 which relates the number of thermodynamic conjugate pairs $\textsf{W}$ and coexistent phases $\textsf{P}=n$ to the degrees of freedom $\textsf{F}$ of the system.
This increase is largely due to the fact that $n-1$ swallowtails are sufficient to support $n$ phases, whereas $n$ inverted swallowtails are necessary to support $n$ stable phases.

\section{charged black holes}\label{sect4}

Adding charge to a black hole stabilizes it by introducing a cusp, and thus a large positive specific heat branch in the Gibbs free energy. Unlike the inclusion of thermodynamic pressure, which admits a positive specific heat branch that intersects $G=0$  \cite{Kubiznak:2016qmn}, the new branch admitted by electric charge intersects $T=0$, which is the black hole extremal limit. 
This branch exists in the same temperature range as the negative specific heat branch and always possesses a smaller Gibbs free energy, making it possible for the black hole to be stable at any temperature in the allowed range. The cusp is signatory of a maximal temperature. This behaviour is comparable to the Reissner–Nordstr\"{o}m black hole \cite{Altamirano:2014tva}.
Due to the presence of charge, the number of stable phases is now $n=\lfloor \frac{N+1}{2} \rfloor$.

\begin{figure}
	\includegraphics[width=0.483\textwidth]{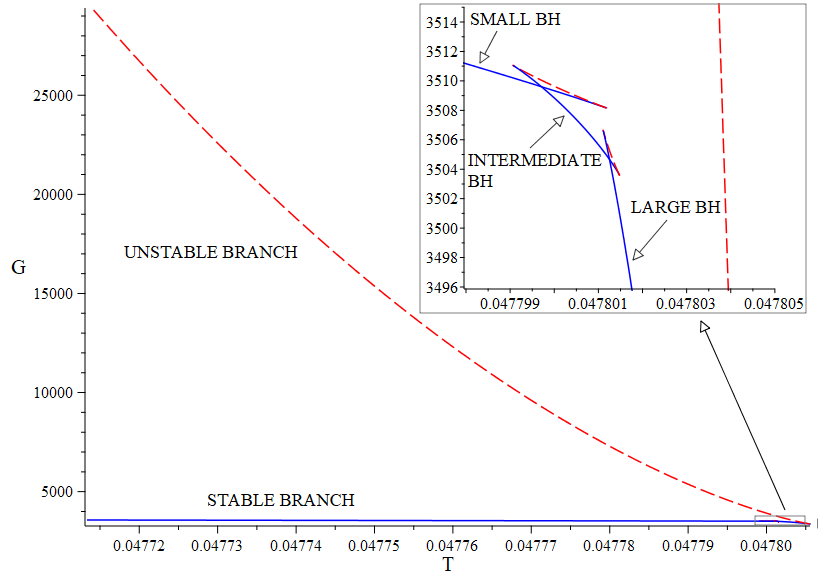}

	\caption{\textbf{$G$-$T$ plot: Two swallowtails on the stable branch of a charged black hole.} Here $d=12$, $N=5$, $\alpha_2 \approx 62.3$, $\alpha_3 \approx 980.8151557$, $\alpha_4 \approx 3858.395052$, $\alpha_5 \approx 2869.228931$, $Q \approx 6.52365409$. Three distinct stable phases separated by two first order phase transitions on the stable branch (blue curves). Dashed curves indicate negative specific heat.}
	\label{fig:5ll-gt-separated} 
\end{figure}

We illustrate these features for a black hole in $d=12$ and $N=5$ in figure~\ref{fig:5ll-gt-separated}.  There are three distinct stable phases,
and first order phase transitions between these stable phases can take place on this branch, indicated by the swallowtails in the inset. These
swallowtails are  fully analogous to those observed in charged AdS black holes.  A triple point can be attained by adjusting $\alpha_2$ so that
the intersection points of the  two swallowtails merge.

\begin{figure}
	\includegraphics[width=0.483\textwidth]{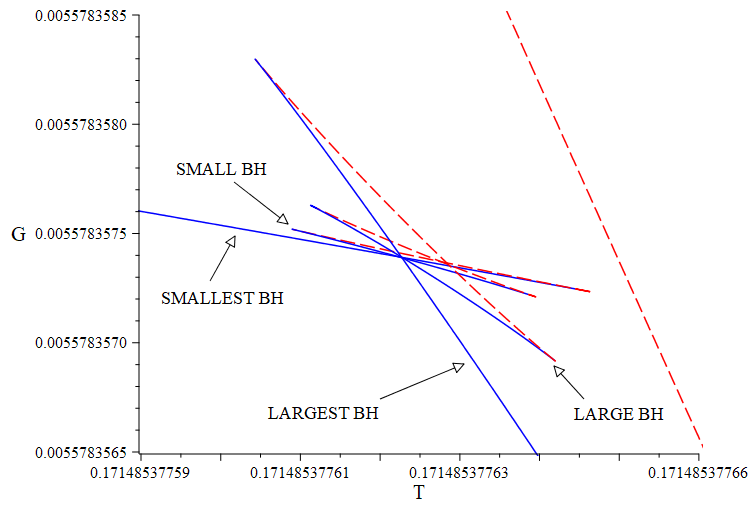}

	\caption{\textbf{$G$-$T$ plot: Stable quadruple point of a charged black hole.} Here $d=16$, $N=7$, $\alpha_2 \approx 10.8404632542$, $\alpha_3 \approx 34.3640023107$, $\alpha_4 \approx 38.0254606761$, $\alpha_5 \approx 15.290149727$, $\alpha_6 \approx 1.99503641246$, $\alpha_7 \approx 0.0566714828054$, $Q \approx 5.16661412$. Four distinct phases (blue curves) coexist at the point where all three swallowtails intersect. Dashed curves indicate negative specific heat.}
	\label{fig:7ll-gt-charged-quad} 
\end{figure}

More swallowtails and stable phases can be introduced by enlarging the value of $N$. Setting $N=7$, in figure~\ref{fig:7ll-gt-charged-quad}, we explicitly present a system with four possible stable states that merge at a quadruple point. As before, $\alpha_2$ takes on a role akin to that of thermodynamic pressure. Increasing $\alpha_2$ separates ordinary swallowtails, but decreasing $\alpha_2$ separates inverted swallowtails, in contrast to the uncharged case.

\begin{figure}
	\includegraphics[width=0.483\textwidth]{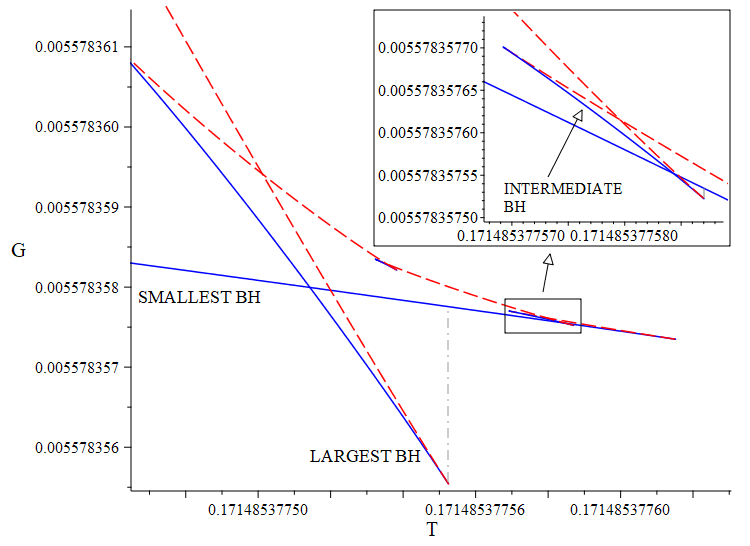}
	\includegraphics[width=0.483\textwidth]{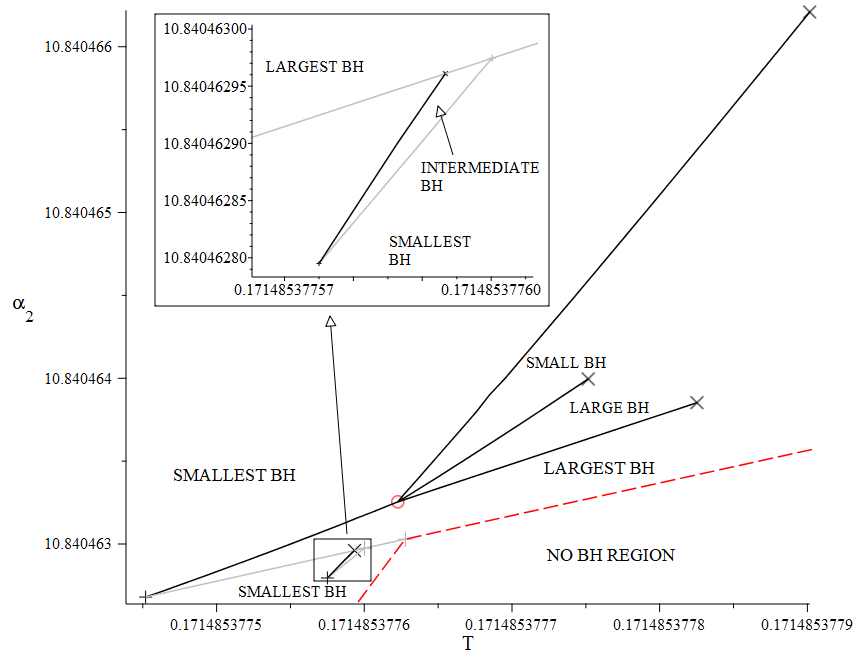}

	\caption{\textbf{$G$-$T$ plot and $\alpha_2$-$T$ phase diagram.} Here $d=16$, $N=7$, $\alpha_3 \approx 34.3640023107$, $\alpha_4 \approx 38.0254606761$, $\alpha_5 \approx 15.290149727$, $\alpha_6 \approx 1.99503641246$, $\alpha_7 \approx 0.0566714828054$, $Q \approx 5.16661412$. {\it Top.} At $\alpha_2 = 10.84046288$, all three swallowtails have been shifted to the unstable branch in the Gibbs free energy. Two inverted swallowtails intersect the stable Gibbs free energy branch and admit three stable phases separated by first order and zeroth order phase transitions. Dashed curves indicate negative specific heat. Dashed gray lines rep-
resent zeroth order phase transitions. {\it Bottom.} The overall phase behaviour is displayed for a range of $\alpha_2$ where phase transitions exist. Black curves represent first order phase transitions, grey curves represent zeroth order phase transitions, and red dashed curved represent the maximal temperature. Black crosses indicate critical points, grey crosses indicate points at which zeroth order phase transitions terminate, and the red circle indicates the quadruple point.}
	\label{fig:7ll-pt} 
\end{figure}

 Increasing $\alpha_2$ above the multi-critical value will separate the swallowtails into multiple stable first order phase transitions at different temperatures, which  diminish in size as $\alpha_2$ increases, and vanish at critical points for sufficiently high values of $\alpha_2$.  The phase diagram is given in the lower diagram of figure~\ref{fig:7ll-pt}. Unlike systems where the Hawking-Page transition is available, there are no phase transitions for very large $\alpha_2$. 
 
 The situation when $\alpha_2$ is decreased below the multi-critical value is a bit different.  For small decreases in  $\alpha_2$, all but one swallowtail becomes embedded in the largest one, and only one stable phase transition exists,  similar to the situation for charged AdS black holes in Einstein gravity.  However, some embedded swallowtails can be shifted to the unstable branch and become inverted swallowtails.  As $\alpha_2$ continues to decrease, this eventually happens to the largest stable swallowtail, and new first order and zeroth order phase transitions emerge as $\alpha_2$ is lowered further. All phase transitions eventually vanish as inverted swallowtails diminish in size for smaller $\alpha_2$. The newly developed first order and zeroth order phase transitions terminate in pairs at the same point, and one such pair of transitions emerges from the zeroth order coexistence curve of another pair. The Gibbs free energy showing the new inverted swallowtails can be found in the upper diagram in figure~\ref{fig:7ll-pt}. The phase diagram for $\alpha_2 > \alpha_{2_{mc}}$ is qualitatively the same at that of the previously discovered multi-critical points for AdS black holes \cite{Tavakoli:2022kmo,Wu:2022bdk,Wu:2022}, barring the maximal temperature.

The maximal temperature is a consequence of the two-branch structure of the Gibbs free energy; neither branch can extend to arbitrarily high temperatures by virtue of the cusp connecting them.  The dashed red line in the phase diagram delineates the maximal temperature, and corresponds to the cusp in $G$.

An outline of a proof for the existence of a maximal temperature can be constructed using limits of $T(r_+)$. Given $d\ge 2N+1$ and $\alpha_k\ge 0\ \forall k$, we  obtain the limits 
\be
\lim_{r_+ \to 0^+} T(r_+) = -\infty, \quad \lim_{r_+ \to \infty} T(r_+) = 0\; .
\ee
$T(r_+)$ is continuous on $(0,\infty)$ if all nonzero Lovelock constants are of the same sign, and we assume $T$ has at least one positive region $\mathbb{I}\in (0,\infty)$ to guarantee physically meaningful behaviour. 

By the intermediate value theorem, $T$ attains at least one root $T(r_+ = a)=0$ between $r_+=0$ and any $r_+ \in \mathbb{I}$. We choose $b$ such that $T(b) \ge \abs{T(r_+)} > 0 \ \forall r_+ \in [a,a+\epsilon] \bigcup [b,\infty), \epsilon >0 $, which must exist as required by the continuity of $T$ and the limit as $r_+ \to \infty$. By the extreme value theorem, $T$ attains its maximum value $T^*=T(R_+)$ on the compact interval $[a+\epsilon, b]$ for some $R_+ \in [a+\epsilon, b]$. Since $[a+\epsilon, b]\bigcap \mathbb{I} \ne \emptyset$, $T^*$ is necessarily positive. With our choice of $b$, we now have $T^* \ge T(b) \ge \abs{T(r_+)} \ \forall r \in (0,a+\epsilon] \bigcup [b,\infty)$, hence $T^*$ is the global maximum on the domain $(0,\infty)$.

By the generalized Gibbs phase rule \cite{Sun:2021gpr}, we find that charged asymptotically Lovelock black holes have $n+1$ thermodynamic degrees of freedom, where $n$ is the number of coexistent phases. 

\section{Conclusions}

We have demonstrated for the first time that asymptotically flat black holes can  
undergo first order phase transitions. 
Thermodynamically such black holes correspond to a system with zero pressure, and so it is somewhat surprising that such transitions can exist. Indeed, they do not exist in Einstein gravity, but   Lovelock gravity  has a sufficiently large number of thermodynamic variables to enable distinct black hole phases.  The number of phases is $n=\lfloor \frac{N-1}{2} \rfloor$, and we have taken the dimension $d$ to be the smallest even number in which the $N$-th order Lovelock term is nontrivial.

The phase behaviour of asymptotically flat black holes is notably different than that of their AdS counterparts. Neutral black holes are generally unstable (having negative specific heat) for most values of their radii, analogous to Schwarzschild black holes. However 
(depending on the parameters) 
there exists a narrow range of temperatures where multiple stable black hole phases exist, with first order transitions between them. Both unstable and stable multicritical points can exist depending on the number and relative values of the thermodynamic parameters.

Charged black holes, by contrast, are thermodynamically stable for a broad range of radii ranging from the $T=0$ extremal case 
up to some maximal value of the temperature $T$. 
These black holes can have stable multiple phases and stable multicritical points. An unstable branch of charged black holes also exists over the same temperature range, and this branch can be punctuated by regions of stable black hole phases that can undergo first order phase transitions.

We have considered only Lovelock theories for which the quadratic curvature coupling parameter $\alpha_2$ is non-vanishing.  This quantity plays a role analogous to pressure in the AdS case. 
Should $\alpha_2 = 0$ we expect that similar behaviour
in $N$-th order Lovelock gravity 
would ensue, with the next largest nonzero $\alpha_k$ playing a similar role. It would be interesting to see how the phase structure is modified in this case.  

Perhaps the most salient lesson that can be drawn from our results is that chemical features of black holes are not limited to AdS spacetimes. Should higher curvature terms of sufficiently large order play a role in quantum gravity, then the phase behaviour of asymptotically flat black holes should emerge as a consequence.

\section*{Acknowledgements}
\label{sc:acknowledgements}

This work is supported in part by the Natural Sciences and Engineering Research Council of Canada (NSERC).  Perimeter Institute and the University of Waterloo are situated on the Haldimand Tract, land that was promised to the Haudenosaunee of the Six Nations of the Grand River, and is within the territory of the Neutral, Anishnawbe, and Haudenosaunee peoples.

\bibliographystyle{JHEP}
%\bibliography{refs}

\providecommand{\href}[2]{#2}\begingroup\raggedright\endgroup

\end{document}